\documentclass[preprint,12pt]{elsarticle}

\usepackage{lineno,hyperref}
\usepackage{amssymb}
\usepackage{amsthm}
\usepackage{amsmath}
\usepackage{epsfig}
\usepackage{mathtools}

\journal{Journal of Magnetism and Magnetic Materials}

\begin{document}

\begin{frontmatter}

\title{On semiconductor--metal transition in FeSi induced by ultrahigh magnetic field}

\author{Yu. B. Kudasov \corref{mycorrespondingauthor}}
\ead{kudasov@ntc.vniief.ru}
\cortext[mycorrespondingauthor]{Corresponding author}

\author{D. A. Maslov}

\address{Sarov Physics and Technology Institute, NRNU "MEPhI", 6 Dukhov st.,Sarov, 607186, Russia}
\address{Russian Federal Nuclear Center - VNIIEF, 37 Mira av., Sarov, 607188, Russia}

\begin{abstract}
At low temperatures, iron monosilicide is a strongly correlated narrow-gap semiconductor. A first order 
transition to metal state induced by magnetic field was observed for the first time at 355~T in Ref.~[Yu. B. 
Kudasov {\em et al}., JETP Lett. {\bf 68} (1998) 350]. However, recently a smooth transition from 230~T to 
270~T was found under similar conditions in Ref.~[D. Nakamura {\em{et al}}., Phys. Rev. Lett. {\bf{127}} 
(2021) 156601]. This discrepancy goes far beyond experimental errors and deserves a careful study. A methodological analysis of inductive and RF techniques of conductivity measurements shows that the 
difference of these critical magnetic field estimations stems from a divergence in dynamic ranges of the 
techniques. In fact, the above mentioned methods supplement each other. The semiconductor-metal transition 
under magnetic field in FeSi is a complex phenomenon which occurs at the wide range of magnetic fields.

\end{abstract}


\end{frontmatter}


\section{Introduction}

Iron monosilicide (FeSi) has been attracting attention of theorists and experimentalists for several decades 
\cite{Jaccarino,Schlesinger,Paschen,Fang,Changdar}. It is a non-magnetic metal above room 
temperature\cite{Jaccarino,Changdar}. However, FeSi becomes a narrow-gap semiconductor below 100~K with the 
gap width of about 60~meV \cite{Schlesinger,Changdar}. Effective masses of mobile charge carriers on the top 
of the valence band and the bottom of the conduction band grow with decreasing temperature and reach extremely 
large values ($m \sim 100 m_0$), which makes it possible to attribute FeSi to a rather exotic group of heavy 
fermion compounds without f-elements \cite{Schlesinger,Fang}.
%
Comparison of experimental data with results of calculations of the electronic structure demonstrated that the 
heavy carriers in FeSi appeared due to strong electronic correlations in bands formed mainly by iron 
d-orbitals \cite{Arita,Tomczak}. This problem was investigated in the framework of the Hubbard model 
\cite{Glushkov1}. The hypothesis of FeSi as a Kondo insulator was widely discussed \cite{Schlesinger97,Klein}.
However, a realistic model should include strong multiorbital correlations \cite{Tomczak}.

Below 70~K another transport anomalies in FeSi are observed. They are ascribed to quasiparticles of 
spin-polaron type, which form a very narrow band about 6~meV wide inside the gap \cite{Glushkov2}. During the 
last decade, few Kondo insulators were proven to be a topological insulator including SmB$_6$, which 
thermodynamical and transport low-temperature properties were similar to those of FeSi 
\cite{Breindel2,Breindel,Kim}. The existence
of a thin layer with high conductivity on the surface of iron monosilicide was recently experimentally 
demonstrated
on ultra-thin single crystal samples \cite{Fang,Ohtsuka}.
Moreover, the observed hysteresis of the Hall effect demonstrated that the 2D conductive surface layer in FeSi 
is ferromagnetic with a non-magnetic bulk state \cite{Ohtsuka}.

At low temperatures and moderate magnetic fields, anisotropic magnetoresistance was revealed in FeSi 
\cite{Fang}. The anisotropy  was a consequence of an interplay of
the surface and bulk charge mobile carriers  \cite{Breindel2} and transport properties of FeSi was described 
by means of two-band model \cite{Paschen}. Under ultrahigh magnetic field (above 100~T), the gap in
electron spectrum of FeSi should be suppressed by the Zeeman splitting of the conduction and valence bands 
edges.
The LDA+U \cite{Anisimov} and LSDA \cite{Kulatov,Arioka} calculations gave an estimation of the critical 
magnetic field of the transition to the metallic state of about 170 T at $T=0$~K. In fact, the mechanism of 
metallization involves a renormalization of the quasiparticle bands by spin fluctuations that  could greatly 
shift this value \cite{Povzner}.

For the first time, the semiconductor-metal transition induced by ultrahigh magnetic field in FeSi was 
experimentally observed and studied in Ref.~\cite{Kudasov1,Kudasov2} using induction magnetization 
measurements and radio-frequency (RF) contactless conductivity measurements. At $T=5$~K it was observed at 355 
T and was accompanied by a step in the magnetic moment. At $T=78$~K there was a smooth increase in 
conductivity, and without the step in the magnetic moment. Qualitatively, this was in good agreement with the 
theoretical magnetic phase diagram \cite{Anisimov}: a phase transition of the first kind from singlet 
semiconductor to ferromagnetic metal existed at low temperatures and a smooth transition did above a certain 
critical temperature. Recently, FeSi was studied under ultrahigh magnetic field at various temperatures using 
a similar RF measuring technique \cite{Nakamura1} and at $T=6$~K an extended transition was observed, which 
ended at about 275 T. The discrepancy in the results of Ref.~\cite{Kudasov1,Kudasov2,Nakamura1} goes far 
beyond experimental errors and deserves a careful study.

\begin{figure}
\includegraphics[width=0.88\textwidth]{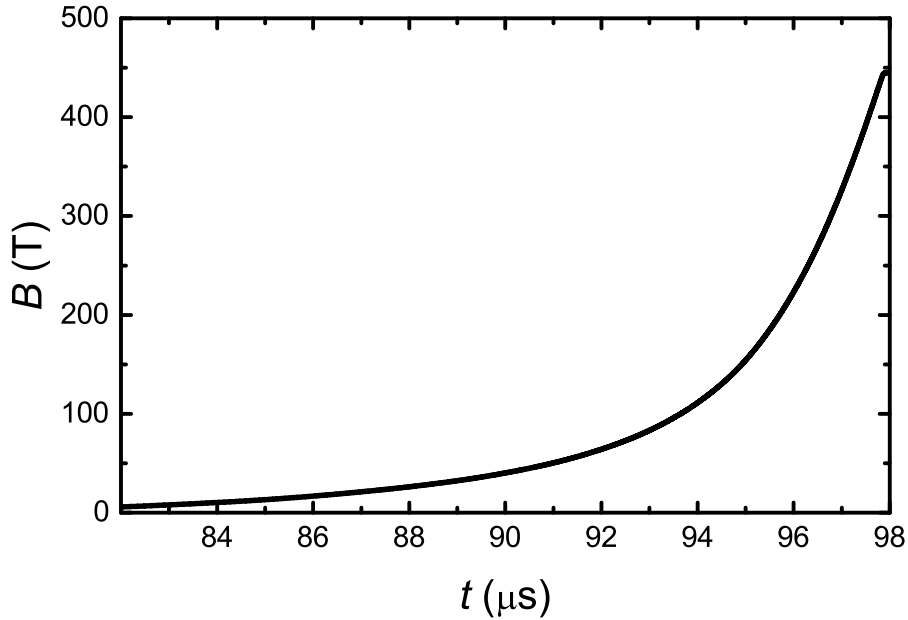}
\caption{\label{f1} The final part of the magnetic field pulse of
MC-1 generator.}
\end{figure}

A magnetic flux compression technique is the only possibility to investigate transport and magnetic properties 
of substances at 300~T and above \cite{Shneerson}. The fast flux compression is realized by means of high 
explosives (HEC) \cite{Boriskov} or ponderomotive electromagnetic forces (EMC) \cite{Nakamura}. The both 
approaches were used in the investigations of FeSi: the first one was implemented in MC-1 generator, which was 
used in Ref.~\cite{Kudasov1,Kudasov2}, the EMC technique was applied in Ref.~\cite{Nakamura1}. A time dependence of 
magnetic field in these devices had a complex shape: a slow initial part up to approximately 16~T and fast 
growth of the magnetic field under the flux compression during about 15~$\mu$s. The final part of the magnetic 
field pulse of the MC-1 generator is shown in Fig.~\ref{f1}. It is very similar to that of EMC facility. It 
showld be mentioned that the ultrahigh magnetic field generation is accompanied by intensive electromagnetic 
noises, and, that is why, a choice of measuring methods is greatly limited \cite{Boriskov}.

In the present article, we consider methodological aspects of inductive and RF conductivity 
measuring techniques and discuss the discrepancy of the results on the semiconductor--metal transition under 
the ultrahigh magnetic field in FeSi.

\section{Inductive conductivity measurement}

\begin{figure}
\centering
\includegraphics[width=0.25\textwidth]{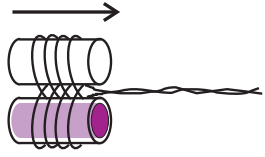}
\caption{\label{f2} Schematic view of compensated inductive sensor. The arrow denotes the external magnetic field direction.}
\end{figure}

\begin{figure}
    \includegraphics[width=0.68\textwidth]{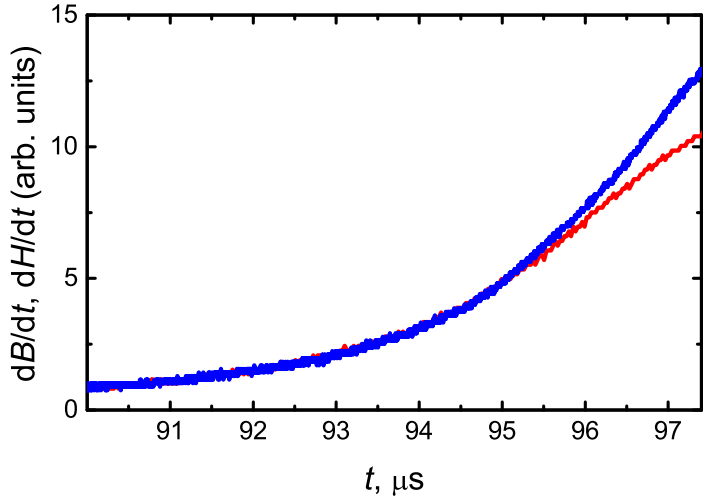}
    \caption{\label{f3} Normalized background signal from pick-up coil (blue line)
    and compensation sensor signal (red line)  \cite{Kudasov1}.}
\end{figure}

The compensation inductive technique is widely used for magnetization and conductivity measurements in
pulsed magnetic fields and, in particular, in ultrahigh magnetic fields \cite{Kudasov1,Kirste}. The 
compensation sensor is a pair of small identical coils with counter-winding, which axes are oriented along the 
external magnetic field (see Fig.~\ref{f2}). A measured sample is installed in one of the coils. In case of a long 
sample, the electromotive force generated by the sensor consists of two components \cite{Kudasov3}:
\begin{eqnarray}
    \mathcal{E}= \alpha \mu_0 \dot{H} + \mu_0 N S \dot{M},
    \label{E}
\end{eqnarray}
where $H$ and $M$ are the magnetic intensity and magnetization (or specific magnetic moment appeared due to 
eddy currents), $\mu_0$ is the permeability of free space, $N$ is the number of turns, $S$ is the sample 
sectional area, and $\alpha$ is the coefficient, which is
proportional to compensation error. From Eq.~(\ref{E}) one can see
that there is an additional background signal which is proportional
to the time derivative of the magnetic flux density. It can be eliminated by comparing it with a background signal obtained from the pick-up coil which is used for magnetic field measurement. The procedure is
illustrated in Fig.~\ref{f3} where signals from Ref.~\cite{Kudasov1}
at $T=77$~K are shown.

In a conductive sample, a pulsed magnetic field generates  eddy
currents, which induce the specific magnetic moment $M$. To separate
the contribution of the sample conductivity to the signal from
intrinsic magnetization, an additional measurement has to be
performed with a powder sample in
dielectric matrix \cite{Kudasov2}. It gives the magnetization
response only.

Under ultrahigh magnetic fields a number of specific requirements to
the sensor have to be taken into account. For example, due to high
rates of magnetic field growth a voltage reaches huge values even in
miniature sensors and can cause an electrical breakdown. To
eliminate the problem a special type of winding can be used
\cite{Kudasov3}.

When conductivity is small, that is, the magnetic field generated by
eddy currents in the sample is much less than the external pulsed
magnetic field, conductivity can be determined from the measured
electromotive force by means of the following expression
\cite{Kudasov2}
\begin{eqnarray}
\sigma(t)=\frac{8}{\pi R^4 \mu_0 \dot{B} N}\int_{0}^{t}\mathcal{E}(\tau) \mbox{d}\tau
\label{sig}
\end{eqnarray}
where $R$ is the sample radius. Here the sample is assumed to be a
long cylinder. The integral form of this equation partially
suppresses an effect of noises.

Eq.~(\ref{sig}) allows estimating a threshold of sensitivity. To do
this, the integral in the right side of the equation should be
replace by its minimal detectable value. Then, the threshold of
sensitivity is inversely proportional to the time derivative of the
magnetic flux density, which varies drastically during the pulse of
the magnetic filed as one can see in Fig.~\ref{f1}. The upper bound
of the dynamical range is defined by a non-linear regime of magnetic
field diffusion into the sample while the conductivity becomes
sufficiently high. In practice, however, it was limited to the
maximum value of the recorded signal \cite{Kudasov2}.

It should be mentioned that an effect of the finite length of the
sample on the measurement results can be taken into account by means
of a magnetostatic reciprocity theorem \cite{Kirste}.

\section{RF conductivity measurement}

The RF method of conductivity measurement has been successfully
applied in ultrahigh magnetic field for a long time
\cite{Miura,Kudasov1,Kudasov2,Kudasov4,Kudasov5,Nakamura1,Nakamura2}.
The absence of contacts with a sample and using a narrow frequency
band makes it possible to suppress electromagnetic
interference that are induced by ultrahigh magnetic field.
Initially, two small flat coils with a typical diameter of 3 to 5~mm
were used, each of which contained several turns. The coils were
positioned coaxially with each other. A
test sample in the form of a plate was inserted into a small gap between the coils \cite{Miura}. The axis of the coils was
oriented perpendicular to the external magnetic field to reduce the
induced voltage. An RF signal from the generator was fed to one of
the coils, and the signal from the other coil was sent to the
oscilloscope. An amplitude of the measured signal depended on the
conductivity of the sample. An analysis of the electrodynamics of
the measuring unit showed that there is a wide area of linear
conductivity response.

\begin{figure}
	\includegraphics[width=0.55\textwidth]{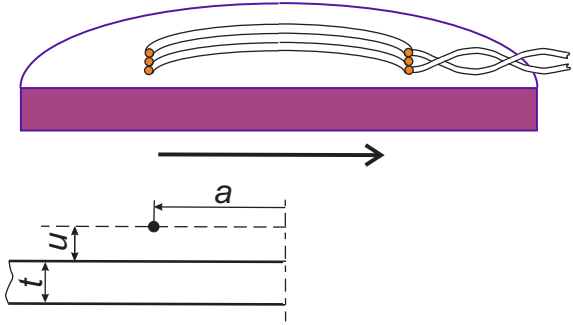}
	\caption{\label{f4} Schematic view of RF single-coil measuring unit (the coil and sample) and model representation (bottom panel). The arrow denotes the external magnetic field direction.}
\end{figure}

Currently, a single-coil measuring unit is used
\cite{Kudasov1,Kudasov4,Nakamura1,Nakamura2} (see Fig.~\ref{f4}). It
combines the function of the transmitting and receiving coils. A
probing RF signal is applied to the coil through a bandpass filter
and cable line. A reflected signal is generated by eddy currents in
the sample. It propagates through the same cable and bandpass
filter. Then the reflected signal is separated from the probe one.
The electrodynamics of RF measurements in the two-coil system was
thoroughly investigated \cite{Miura}. Below we will consider a
single-coil system using the same approach.

The RF measuring unit has an axial symmetry (Fig.~\ref{f4}).
The vector potential in
cylindrical coordinates ($r$, $z$, $\varphi$) is determined by the
following expression \cite{Miura}
\begin{eqnarray}
\frac{\partial^2 A}{\partial r^2}+\frac{1}{r}\frac{\partial A}{\partial r}-\frac{A}{r^2}+\frac{\partial^2 
A}{\partial z^2}=i\omega \mu_0 \sigma A,
\label{A}
\end{eqnarray}
where $A \equiv A_{\varphi}$ is the $\varphi$-component of the
vector potential, $\omega$ is the circular frequency of the probe
signal.

By separating the variables $r$ and $z$ the Eq.~(\ref{A}) is reduced to Bessel's equation of the first kind on $r$ and the following equation on $z$
\begin{eqnarray}
\frac{\partial^2 A}{\partial z^2}={k^\prime} ^2 A,
\label{Bessel}
\end{eqnarray}
where ${k^\prime}^2 = k^2 + i \omega \mu_0 \sigma$, and $k$ is the separation parameter. Then, a solution of Eq.~\ref{A} under the physical boundary condition at the $z$-axis can be written as
\begin{eqnarray}
A(r,z)=\int_0 ^{\infty} \big[a(k)e^{k^\prime z}+b(k)e^{-k^\prime z} \big] J_1 (kr) \mbox{d} k,
\label{general}
\end{eqnarray}
where $a(k)$ and $b(k)$ are the coefficients, $J_1 (kr)$ is the Bessel function of the first kind. The whole
space is separated into four regions as shown in the bottom panel
of Fig.~\ref{f4}. The coefficients $a(k)$ and $b(k)$ in each of them
are determined by the boundary conditions
\cite{Miura}. There is a significant difference with the two-coil
system. Namely, the total voltage on the coil is divergent as it is
on the transmitting coil in the two-coil system. This is an effect
of the approximation of infinitely thin coil. That is why, instead
of an amplitude of the total signal we obtain a normalized amplitude of the
reflected signal induced by the eddy currents. It has the following
complex form
\begin{eqnarray}
R=\frac{U}{U_\infty}=
2 \int\limits_0 ^\infty { \frac{({k^\prime}^2 -k^2) \exp(-2ku) \sinh (k^\prime t) J_1
^2(ka)} {(k+  k^\prime)^2 \exp(k^\prime t) - ({k^\prime} -k)^2 \exp(-k^\prime t)}} \mbox{d} k   \nonumber \\ \times
\Bigg[\int\limits_0 ^\infty \exp(-2ku) J_1 ^2(ka) \mbox{d}k \Bigg]^{-1}, \label{R}
\end{eqnarray}
\noindent where $a$, $t$, and $u$ are the coil radius, sample
thickness, and width of the gap between the coil and sample Fig.~\ref{f4}, $U_\infty$ is the amplitude of the reflected signal at $\sigma\rightarrow \infty$.

\begin{figure}
\includegraphics[width=0.68\textwidth]{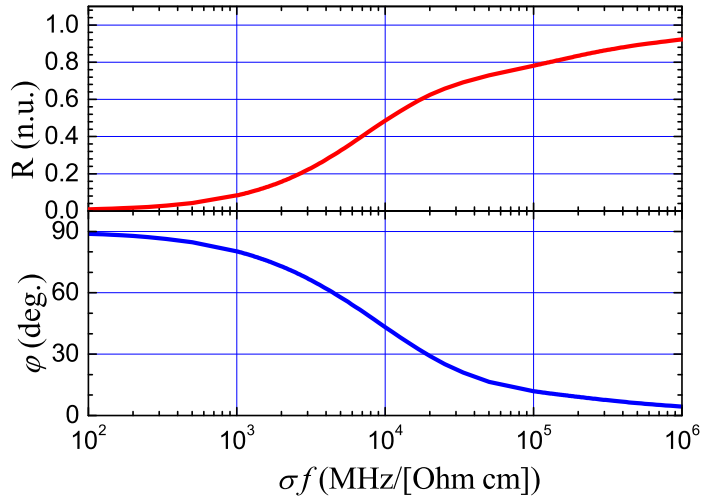}
\caption{\label{f5} Amplitude $R$ and phase $\varphi$ of the normalized
reflected signal of the RF single-coil measuring unit (Eq.~\ref{R})
as a function of sample conductivity and frequency of the probe
signal. Parameters correspond to Ref.~\cite{Kudasov2}: $a=1.5$~mm,
$t=0.3$~mm, $u=0.2$~mm (see Fig.~\ref{f4}).}.
\end{figure}

An amplitude and phase of the reflected signal are shown in Fig.~\ref{f5}. The geometrical parameters
correspond to those of measuring unit in Ref.\cite{Kudasov2}. Since
the amplitude and phase of the reflected signal depend on the product $\sigma f$, the
sensitivity threshold of conductivity inversely proportional to the
frequency $f$. Under heavy electromagnetic interferences the dynamical
range of the technique is usually no greater than two orders of
magnitude of conductivity.

The analysis above gives a good description for the RF measurements
at relatively low frequency band ($f \approx 50$~MHz)
\cite{Miura,Kudasov2,Kudasov4}. At the same time, while high
frequency is applied ($f \approx 500\div700$~MHz) \cite{Nakamura1},
a role of interturn and coil-sample capacitances increases.
Therefore, the coil becomes a resonant circuit, and the analysis of
the measuring unit operation becomes more complex although the underlying physical phenomenon remains the same.
The operation of high-frequency measuring unit was studied in
detail in Ref.~\cite{Nakamura2}.

\section{Semiconductor-metal transition in FeSi}

\begin{figure}
\includegraphics[width=0.68\textwidth]{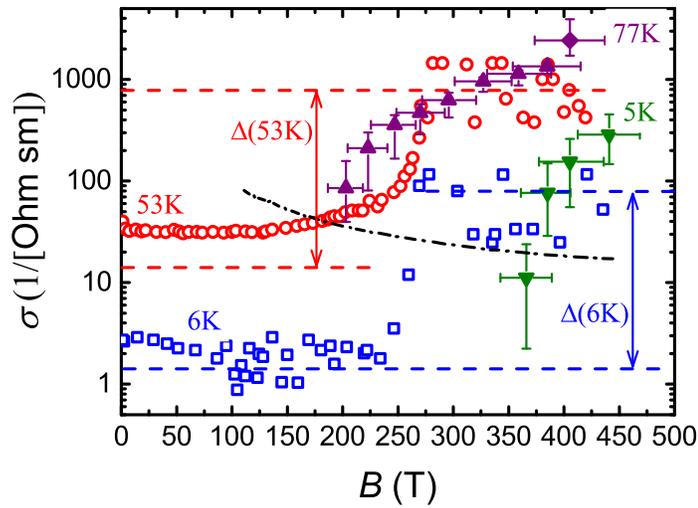}
\caption{\label{f6} Magnetic field dependence of conductivity of
FeSi at different initial temperatures. The full and open symbols
correspond to the results of Ref.~\cite{Kudasov2} and
\cite{Nakamura1}, correspondingly. The dash lines are estimations of
the dynamical ranges $\Delta(T)$ \cite{Nakamura1,Nakamura2}, the dash-dot black
line is the low threshold of the inductive technique sensitivity
(see Eq.~\ref{f2}). The circles and squares are RF measuremnts at
700~MHz \cite{Nakamura1}, the diamond is that at 50~MHz
\cite{Kudasov2}, the up- and down-triangles demonstrate the results
of inductive measurements \cite{Kudasov2}}.
\end{figure}

Experimental data from Ref.~\cite{Kudasov2} and \cite{Nakamura1} are
shown in Fig.~\ref{f6} together with estimations of dynamical ranges
corresponding to different techniques. One can see that they are in
a reasonable agreement with each other. The RF technique at 700~MHz
\cite{Nakamura1} was sensitive to much lower conductivity than that
at 50~MHz and even more so the inductive compensation method.
That is why, it gave the magnetic field dependence of conductivity
starting from the initial values up to about 10$^2$~[Ohm cm]$^{-1}$
at 6~K and 10$^3$~[Ohm cm]$^{-1}$ at 53~K. The inductive technique
had the much higher sensitivity threshold and provided measurement
above these values that occurred at higher magnetic field. One can
see in Fig.~\ref{f6} that the low-temperature field dependencies of conductivity of
Ref.~\cite{Kudasov2,Nakamura1} are in a reasonable agreement and
supplement each other.

On the other hand, the results in Fig.~\ref{f6} demonstrate that the
semiconductor-metal transition in FeSi is a complex phenomenon which
is extended in magnetic field \cite{Kudasov2,Nakamura1}.
Magnetoresistance of FeSi in moderate magnetic field up to 10~T is
determined by the interplay of the surface and bulk conductivity
\cite{Fang,Paschen,Breindel2}. At higher magnetic field the in-gap
spin-polaron-like states should play an important role because their
characteristic energy scale is about 10~meV \cite{Glushkov2}. Then,
the collapse of the intrinsic band gap occurs at ultrahigh magnetic
fields. The first order transition with the magnetization step was
observed at 5~K, and it was absent at 77~K \cite{Kudasov2}. The line
of the first order transition with the critical point
in the magnetic phase diagram \cite{Anisimov} is still an open
question which can be resolved by magnetization measurements at
intermediate temperatures.

\section{Acknowledgement}

The work was supported by National Center Physics and Mathematics
(project "Investigation under high and ultrahigh magnetic fields"). The
authors would like to thank Professor A.E. Dubinov  for valuable
comments.

\section*{References}

\bibliography{bibFeSi}

\end{document}